\pdfoutput=1

\documentclass[11pt,a4paper]{article}

\usepackage{jheppub}
\usepackage{amsmath}
\usepackage{multicol,bbm}
\usepackage{enumerate}
\usepackage{bibentry}

\usepackage{graphicx}
\usepackage{dcolumn}
\usepackage{float}
\usepackage[many]{tcolorbox}
\usepackage{lipsum}
\usepackage{cancel}

\usepackage{shuffle}
\usepackage{graphbox}
\usepackage{environ}
\usepackage{physics}

\usepackage{amssymb}
\usepackage{amsthm}
\newtheorem{theorem}{Theorem}
\usepackage{bm}
\usepackage{multirow}
\usepackage{caption}
\usepackage{tabularx}
\usepackage{enumitem}

\usepackage{extarrows}
\usepackage{tikz}
\usepackage{verbatim} 

\def\be{\begin{equation}}
\def\ee{\end{equation}}
\def\ba{\begin{eqnarray}}
\def\ea{\end{eqnarray}}

\def\CP1{\mathbb{CP}^1}
\def\SL2C{\mathrm{SL}(2,\mathbb{C})}

\def\Z2{\mathbb{Z}_2}

\def\su2{{SU(2)}}

\def\[{\left[}
\def\]{\right]}

\def\({\left(}
\def\){\right)}
\def\[{\left[}
\def\]{\right]}

\def\<{\langle}
\def\>{\rangle}

\def\i2{\frac{i}{2}}

\def\2F1{\,_2{\rm F}_1}

\graphicspath{{figures/}}

\begin{document}


\title{Emergent unitarity, all-loop cuts and integrations from the ABJM amplituhedron}


\date{\today}
\author[a,b,c,d]{Song He,}
\author[e]{Chia{-}Kai Kuo,} 
\author[f]{Zhenjie Li,}
 \author[a,g]{Yao{-}Qi Zhang}%
\affiliation[a]{CAS Key Laboratory of Theoretical Physics, Institute of Theoretical Physics, Chinese Academy of Sciences, Beijing 100190, China}
\affiliation[b]{
School of Fundamental Physics and Mathematical Sciences, Hangzhou Institute for Advanced Study, UCAS, Hangzhou 310024, China}
\affiliation[c]{ICTP{-}AP
International Centre for Theoretical Physics Asia{-}Pacific, Beijing/Hangzhou, China}
\affiliation[d]{Peng Huanwu Center for Fundamental Theory, Hefei, Anhui 230026, P. R. China}
\affiliation[e]{Department of Physics and Center for Theoretical Physics, National Taiwan University, Taipei 10617, Taiwan}
\affiliation[f]{SLAC National Accelerator Laboratory, Stanford University, Stanford, CA 94309, USA}
\affiliation[g]{School of Physical Sciences, University of Chinese Academy of Sciences, No.19A Yuquan Road, Beijing 100049, China}
\emailAdd{songhe@itp.ac.cn}
\emailAdd{chiakaikuo@gmail.com}
\emailAdd{zhenjiel@slac.stanford.edu}
\emailAdd{zhangyaoqi@itp.ac.cn}

\preprint{ \begin{flushright} USTC-ICTS/PCFT-23-07\\ SLAC-PUB-17725\end{flushright}}

\date{\today}

\abstract{We elaborate on aspects of a new positive geometry proposed recently, which was conjectured to be the four-point amplituhedron for ABJM theory. We study generalized unitarity cuts from the geometry, and in particular we prove that (1) the four-point integrand satisfies perturbative unitarity (or optical theorem) to all loops, which follows directly from the geometry, and (2) vanishing cuts involving odd-point amplitudes follow from the ``bipartite" nature of the associated ``negative geometries", which justifies their appearance in ABJM theory. We also take a first step in integrating the forms of these negative geometries and obtain an infrared-finite quantity up to two loops, from which we extract the cusp anomalous dimension at leading order.}


\maketitle

\section{Introduction and review}\label{sec1}

The amplituhedron of planar ${\cal N}=4$ super Yang-Mills theory (SYM)~\cite{Arkani-Hamed:2013jha,Arkani-Hamed:2013kca, Arkani-Hamed:2017vfh} is a surprising geometric structure, where the canonical forms of these positive geometries~\cite{Arkani-Hamed:2017tmz} encode all-loop, all-multiplicity scattering amplitudes in the theory, and it has triggered a lot of progress in search for such positive geometries in other theories and contexts ({\it c.f.} \cite{Arkani-Hamed:2017fdk, Arkani-Hamed:2017mur, Arkani-Hamed:2019mrd, Arkani-Hamed:2019vag, Arkani-Hamed:2019plo, Damgaard:2019ztj, Huang:2021jlh, He:2021llb}). In~\cite{He:2022cup}, a new positive geometry was proposed by projecting (or reducing) external and loop momenta to $D=3$ of the four-point amplituhedron in ${\cal N}=4$ SYM. Remarkably, the canonical form of this new geometry was argued to give $L$-loop four-point integrands in ${\cal N}=6$ Chern-Simons-matter theory, or ABJM theory~\cite{Hosomichi:2008jb, Aharony:2008ug}. We have shown that the geometry makes various all-loop cuts, such as soft cuts and vanishing triple cuts, manifest, and we have used it for explicitly computing four-point integrands up to $L=5$~\cite{He:2022cup}, which has confirmed a conjecture at $L=3$~\cite{Bianchi:2014iia} and provided new results for $L=4,5$. 

Independent of the interpretation as ABJM (four-point) amplituhedron \cite{Huang:2021jlh,He:2021llb}, this new geometry has provided a simplified model with rich structures for the amplituhedron in ${\cal N}=4$ SYM. This has become particularly clear when the latter is decomposed into the so-called {\it negative geometries}~\cite{Arkani-Hamed:2021iya}, which can be viewed as natural building blocks for multi-loop amplitudes in ${\cal N}=4$ SYM. As shown in~\cite{He:2022cup}, the reduction to $D=3$ has simplified such geometries enormously: only those negative geometries with corresponding to bipartite graphs survive the reduction; thus going down to $D=3$ not only drastically reduce the number of possible topologies of $L$-loop integrands, but also put very strong constraints on the pole structure of each geometry. This allows us to determine the canonical form of this $D=3$ amplituhedron to $L=5$ without much work, which in turn gives loop integrands for four-point ABJM amplitudes. 

In this note, we further study this new geometry along two directions: deriving some (all-loop) cuts of the loop integrands and computing certain integrated results by performing loop integrations. In the first direction, similar to those done in~\cite{Arkani-Hamed:2013kca, Arkani-Hamed:2018rsk, YelleshpurSrikant:2019meu} for ${\cal N}=4$ SYM, there are numerous generalized unitarity cuts that can be derived from the geometry: as we will see even the simplest ones such as next-to-ladder cuts and their generalizations are already difficult to compute from Feynman diagrams, thus they provide valuable new data for ABJM integrands with more loops and legs (some progress has been made in understanding higher-point ABJM amplituhedron~\cite{progress1}, see~\cite{Huang:2021jlh, He:2021llb} for tree amplituhedron). In section~\ref{sec:uni}, we will first prove that the ABJM four-point amplituhedron satisfies perturbative unitarity (optical theorem) to all loops in a recursive way which is another example of unitarity from positive geometry. The proof of perturbative unitarity for (four-point) amplitudes in $D=4$ (SYM) and the reduced $D=3$ (ABJM) cases are very similar, which shows how ``rigid" we have unitarity encoded in these geometries underlying scattering amplitudes. Then we will focus on an infinite class of cuts special to ABJM, which vanish due to the existence of vanishing odd-particle amplitudes. It is highly non-trivial to see such vanishing cuts from the geometry, which in principle requires cancellation of numerous contributions. We will show, however, that it is exactly the bipartite nature of the geometry that guarantees such cuts to vanish, which strongly supports our conjecture and shows that``bipartite geometries" are destined to describe ABJM amplitudes for all multiplicities. 

On the other hand, a natural question after obtaining loop integrands (given by canonical forms for the geometries) is how to integrate them. It has been shown that by integrating all but one loop variable of the logarithm of amplitudes, one obtains an integrated, infrared-finite, quantity of a single variable depending on the last loop variable; This can be done for individual negative geometries, and have been computed for four-point and five-point amplitudes in ${\cal N}=4$ SYM~\cite{Arkani-Hamed:2021iya, Chicherin:2022bov, Chicherin:2022zxo}. In section~\ref{sec:integration}, we will consider how to perform such integrations for negative geometries for the logarithm of four-point ABJM amplitude. The main result is a computation of the infrared-finite function $F_{L-1}(z)$ for $L\leq 3$ (with different parity property for even and odd $L$) and the confirmation that $L=2$ result gives the leading contribution to cusp anomalous dimension, weighted by a factor of  the 't Hooft coupling $(N/k)^L$ (while $L=1,3$ gives vanishing contributions). As a byproduct, we will also give a reduction identity which allows us to trivially integrate any loop corresponding to a ``leaf'' (valency-1 node) thus reducing certain higher-loop negative geometries to lower-loop ones.

\subsection{Review of ABJM four-point amplituhedron and negative geometries} 

Recall that the $n$-point amplituhedron is defined in the space of $n$ momentum twistors~\cite{Hodges:2009hk},
$Z_a^I$ with $a=1,2,\dots, n$ for external kinematics, as well as $L$ lines in the twistor space, $(AB)^{I J}_i$ with $i=1, \dots, L$ for loop momenta; here $I,J=1,\dots, 4$ are $\operatorname{SL}(4)$ indices, and the simplest bosonic $\operatorname{SL}(4)$ invariant is defined as $\langle a b c d\rangle\equiv \epsilon_{I J K L}Z_a^I Z_b^J Z_c^K Z_d^L$ (and similarly for $\langle (AB)_i a b\rangle$ and $\langle (AB)_i (AB)_j\rangle$). In~\cite{Elvang:2014fja}, external kinematics in $D=3$ was defined by dimensionally reducing every external line, $(Z_a Z_{a{+}1})$; in a completely analogous manner, here we also need to dimensionally reduce all loop variables $(AB)_i$, both of which are achieved by the so-called {\it symplectic conditions} on these lines:
\begin{equation}\label{sympletic}
{\bf \Omega}_{IJ} Z_a^I Z_{a{+}1}^J={\bf \Omega}_{IJ} A_i^I B_i^J=0\,, {\rm with}~{\bf \Omega}=\mqty(0 & \epsilon_{2\times 2} \\ \epsilon_{2\times2} & 0),
\end{equation}
for $a=1, 2, \dots, n$ and $i=1, \dots, L$, where the totally antisymmetric matrix is defined as $\epsilon_{2\times 2}=\mqty(0 & 1 \\ -1 & 0)$. Focusing on four-point case, the reduced $L$-loop $n=4$ amplituhedron in $D=3$ becomes a $3L$-dimensional geometry in $\ell_i\equiv(AB)_{i=1,\dots, L}$ variables. An important subtlety is that $\langle 1234\rangle<0$ for real $Z$'s satisfying symplectic conditions, thus we need to flip the overall sign for the definition of the $D=4$ amplituhedron~\cite{Arkani-Hamed:2013jha}: we require 
\begin{equation}
\langle AB 12\rangle, \langle AB 23\rangle, \langle AB 34\rangle, \langle AB 14\rangle<0, \langle AB 13\rangle, \langle AB 24\rangle>0, \langle (A B)_i (AB)_j\rangle<0,  
\end{equation}
all defined on the support of \eqref{sympletic}.

A convenient parametrization is~\cite{Arkani-Hamed:2013kca}
\begin{equation}\label{eq:parAB}
 (A B)_i=(Z_1+ x_i Z_2- w_i Z_4, y_i Z_2+ Z_3+z_i Z_4)   ,
\end{equation}
and the symplectic condition on $(AB)_i$ becomes $x_i z_i+ y_i w_i-1=0$; the $n=4$ geometry is defined by ($x_{i,j}:=x_i- x_j$ {\it etc.})
\begin{align}\label{def1}
&\forall i: x_i, y_i, z_i, w_i>0, \quad x_i z_i+y_i w_i=1,\nonumber\\
&\forall i,j: x_{i,j} z_{i,j} + y_{i,j} w_{i,j}<0.
\end{align}
We denote this geometry as ${\cal A}_L$ with the canonical form $\Omega({\cal A}_L):=\Omega_L$, which we claim to give $L$-loop planar integrand for four-point ABJM amplitudes (after stripping off the overall tree amplitude). 

Also, it is often useful to convert these expressions back to dual variables ${\bf x}_i \in \mathbb{R}^3$, defined through ${\bf p}_i={\bf x}_{i+1}-{\bf x}_i\equiv {\bf x}_{i, i+1}$ \cite{Chen:2011vv}. Dual conformal invariance becomes manifest by embedding ${\bf x}_i$ in embedding space, i.e., a projective plane in 5 dimensions $X_i=\left(\frac{1}{2} {\bf x}_i^2, 1, \vec{\bf x}_i\right)$, and defining the inner-product as 
\begin{equation}
(i \cdot j):=-2X_i \cdot X_j={\bf x}_{i, j}^2.
\end{equation}
We can convert expressions of momentum twistors above to such inner products, $\langle AB i-1 i\rangle\equiv \langle \ell i-1 i \rangle\propto (\ell\cdot i)$ and $\langle\ell_i\ell_j\rangle\propto(\ell_i\cdot\ell_j)$, with prefactors cancelled in DCI quantities.

In~\cite{Arkani-Hamed:2021iya}, a nice rewriting for the $n=4$ amplituhedron~\cite{Arkani-Hamed:2013kca} was proposed, where it is decomposed into a sum of negative geometries given by ``mutual negativity" conditions, which trivially carries over to our ${\cal A}_L$ in $D=3$; each negative geometry is represented by a labelled graph with $L$ nodes and $E$ edges (edge $(i j)$ for $\langle (AB)_i (AB)_j\rangle>0$ since we reversed all signs, and no condition otherwise), with an overall sign factor $(-)^E$. We sum over all graphs with $L$ nodes without $2$-cycles, 
\begin{equation}
{\cal A}_L=\sum_g (-)^{E(g)} {\cal A}(g),
\end{equation} 
where ${\cal A}(g)$ is the (oriented) geometry for graph $g$. It suffices to consider all {\it connected} graphs, whose (signed) sum gives the geometry for the logarithm of amplitudes~\cite{Arkani-Hamed:2021iya}. Such a  decomposition is useful since each $A_g$ is simpler, whose canonical form is easier to compute. The loop integrand or canonical form for $L=2,3$ reads:
\begin{equation}
    \begin{gathered}
        \centering
    \includegraphics[scale=0.9]{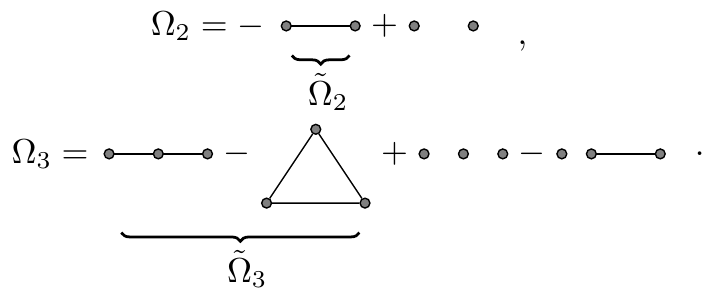}
    \end{gathered}
\end{equation}
where the connected part, or the integrand for logarithm of the amplitude, is denoted as $\tilde{\Omega}_L$, {\it e.g.} $\tilde{\Omega}_2:=\Omega_2-\frac 1 2 \Omega_1^2$. Similarly, the connected part of $L=4$ is given by the sum of graphs with $6$ topologies (and so on for higher $L$),
\begin{equation}
    \includegraphics{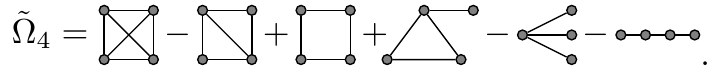}
\end{equation}

What is new in $D=3$ is that most of these geometries do not contribute at all: we find that remarkably, under dimensional reduction only those negative geometries with {\it bipartite} graphs survive in the decomposition. For example, for $\tilde{\Omega}_3$, the chain graph contributes but the triangle does not, {\it i.e.}
\begin{equation}\label{eq:3loopg}
    \underline{\tilde{\Omega}}_3=
\includegraphics[align=c,scale=1.2]{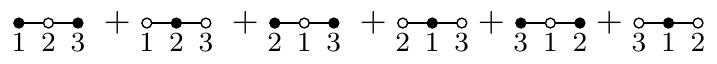}.
\end{equation}

For $\tilde{\Omega}_4$, only the two kinds of tree graphs and the box contribute. This represents a major simplification as the fraction of bipartite graphs in all graphs tends to zero quickly as $L$ increases: for $L=2,\dots, 7$, the number of topologies for connected graphs are $1,2,6,21,112, 853$, but that of bipartite topologies decrease to $1,1,3, 5, 17, 44$, {\it e.g.} for $\tilde{\Omega}_5$, only $5$ topologies (out of $21$) survive the reduction.

Moreover, it turns out that one can compute the canonical form for geometries of bipartite graphs with relative ease, mainly due to their remarkably simple pole structures. We can associate a source/sink vertex (black/white node) for bipartite graphs, and it can only have $s=\langle AB 12\rangle\langle AB34\rangle  \propto y w$ pole, or $t=\langle AB 23\rangle\langle AB14\rangle\propto x z$ pole, respectively. Also there is the mutual pole $D_{i,j}=-\langle\ell_i\ell_j\rangle$ for each link $i-j$:
\begin{equation}
    \begin{gathered}
        \includegraphics[scale=1.2]{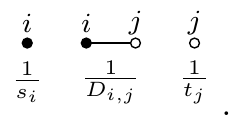}
    \end{gathered}
\end{equation}

All forms of negative geometries up to $L=5$ have been computed~\cite{He:2021llb}, and similar to the $\mathcal{N}=4$ SYM case~\cite{Arkani-Hamed:2021iya}, the forms of all tree graphs can be determined as follows. When attaching node $j$ to $i$ to construct a $L$-node tree, where node $i$ has valency $v_i$ in the original $(L{-}1)$-node tree, all we need is an ``inverse-soft factor" ${\cal T}_{j\to i}$:

\begin{equation}
    \includegraphics[align=c]{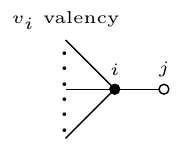}=
    \begin{cases}
    &\frac{2\epsilon_i}{D_{i,j}t_j},\quad  v_i \text{ odd }\\
    &\frac{2c t_i}{\epsilon_iD_{i,j}t_j}, \quad v_i \text{ even}
    \end{cases},
    \includegraphics[align=c]{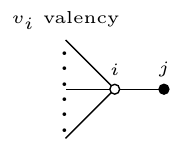}=
    \begin{cases}
    &\frac{2\epsilon_i}{D_{i,j}s_j}, \quad v_i \text{ odd}\\
    &\frac{2c s_i}{\epsilon_iD_{i,j}s_j}, \quad v_i \text{ even}
    \end{cases},
\end{equation}
where we also define the factor $c:=\langle1234\rangle=(2\cdot 4)=-(1\cdot 3)$ and $\epsilon_i:=(\langle 1234\rangle\langle \ell_i13\rangle \langle \ell_i24\rangle)^{1/2}$; equivalently, the $\epsilon$ numerator can be written in embedding formalism as $\epsilon(\ell,1,2,3,4)=\epsilon_{\lambda\mu\nu\rho\sigma}X_l^\lambda X_1^\mu X_2^\nu X_3^\rho X_4^\sigma$, where $X_l^\lambda$ is the loop variable. According to this rule, the $L$-node tree form can be written as the $(L{-}1)$-node tree form times  ${\cal T}_{j\to i}$:
\begin{equation}
\underline{\Omega}_L^{\rm tree} (j \to i)=\underline{\Omega}_{L{-}1} ^{\rm tree}\times {\cal T}_{j \to i}.
\end{equation}

The simplest case is the chain graph which suffices for $L=2,3$. At $L=2$, we have
\begin{equation}
\includegraphics[align=c,scale=1.3]{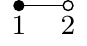}=\frac{2 c^2}{s_1 t_2D_{1,2}}=-\frac{2(1\cdot 3)(2\cdot 4)}{(\ell_1\cdot 2)(\ell_1\cdot 4)(\ell_2\cdot 1)(\ell_2\cdot 3) (\ell_1\cdot \ell_2)}.
\end{equation}
And at $L=3$, \eqref{eq:3loopg} means (note the 2 permutations denote $\ell_2 \leftrightarrow \ell_1, \ell_3$):
\begin{align}
    \underline{\tilde{\Omega}}_3&=\frac{4 c^2 \epsilon_2 }{s_1 t_2 s_3 D_{1,2} D_{2,3}}+ (s\leftrightarrow t) +2~{\rm perms.} \\\nonumber
    &=\frac{4 (2\cdot 4) \epsilon(\ell_2,1,2,3,4) }{(\ell_1\cdot 2)(\ell_1\cdot 4)(\ell_2\cdot 1)(\ell_2\cdot 3) (\ell_3\cdot 2)(\ell_3\cdot 4)(\ell_1\cdot \ell_2)(\ell_2\cdot \ell_3)}+ (13\leftrightarrow 24) +2~{\rm perms.}
\end{align}

For latter convenience, we also record the form for $L$-loop star graph:
\begin{equation}\label{eq:starf}
    \includegraphics[align=c,scale=1.5]{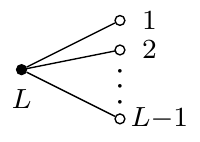}=\frac{2^{L-1}c^{L/2+1}}{s_L \prod_{j=1}^{L-1} t_j D_{j,L}}
    \begin{cases}
        &t_L^{L/2-1},\qquad \text{$L$ is even}\\
        &t_L^{(L-1)/2-1}\epsilon_L c^{\frac{1}{2}},\qquad \text{$L$ is odd}
    \end{cases}.
\end{equation}
For example,for $L=4,5$ we have:
\begin{align}
    \includegraphics[align=c,scale=1.2]{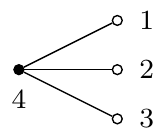}&=
\frac{8 c^3t_4} {s_4 t_1 t_2 t_3 D_{1,4} D_{2,4} D_{3,4} },\\\nonumber
    \includegraphics[align=c,scale=1.2]{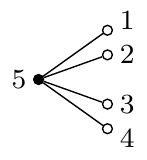}&=\frac{16c^3\epsilon_5  t_5}{s_5t_1t_2t_3t_4D_{1,5}D_{2,5}D_{3,5}D_{4,5}}.
\end{align}

Finally, recall that in ${\cal N}=4$ SYM~\cite{Arkani-Hamed:2021iya}, one can integrate all but one loops for any negative geometry that correspond to a connected graph (thus contribute to the log of amplitude) without having any divergence; doing this for the full integrand for the log of amplitude,  $\underline{\tilde{\Omega}}_L$, gives an important infrared-finite quantity which is related to the Wilson loop with a Lagrangian insertion~\cite{Engelund:2011fg, Alday:2013ip, Chicherin:2022bov, Chicherin:2022zxo}. For ABJM four-point amplituhedron, similarly we define the finite quantity as
\begin{equation}
{\cal W}_L(\ell_1, 1,2,3,4):=\int \prod_{i=2}^L d^3 \ell_i \, \underline{\tilde{\Omega}}_L,  
\end{equation}
which depends on the last loop $\ell_1$ and external points; after stripping off a prefactor it becomes function of a single cross-ratio (see below). This will be the main target of loop integrations for ABJM amplituhedron, and one can extract the ABJM cusp anomalous dimension~\cite{Gromov:2008qe,Griguolo:2012iq,Bianchi:2014ada} by performing the  (divergent) integral over the last loop.


\section{Generalized unitarity cuts to all loops from the geometry}

Essentially the complete information about loop integrands, or canonical forms of the amplituhedron, is encoded in generalized unitarity cuts to all loops, which can be extracted by considering the boundaries of the geometry. In practice we usually compute forms of their boundaries first, which allows us to constrain and even determine the full integrand (as we have seen in~\cite{He:2022cup} up to $L=5$); even for four-point ABJM amplituhedron, one can extract numerous all-loop cuts similar to those for $\mathcal{N}=4$ SYM amplituhedron \cite{Arkani-Hamed:2018rsk, Langer:2019iuo}, which provide stringent consistency checks of the geometry and at the same time predictions for all-loop amplitudes. 

Before proceeding, let us mention a few infinite classes of simple cuts where we only cut $x_i, y_i, z_i, w_i=0$ but not any $D_{i,j}=0$. The simplest example is the so-called ladder cut with $z_1=z_2=\cdots=z_L=0$, and next-to-ladder cuts, where we change some $z$'s to $x$'s, {\it e.g.} $z_1=z_2=\cdots=z_{L-1}= x_L=0$. These are one of the first cuts studied for $\mathcal{N}=4$ SYM amplituhedron~\cite{Arkani-Hamed:2013jha}). Now from the ABJM amplituhedron geometry, it is clear that for any such (next-to-) ladder cuts, only the $L$-point disconnected graph contributes, and the result is
\begin{equation}
\prod_{i\in T} \frac{1}{x_i y_i} \prod_{j \in B} \frac{1}{z_j w_j}\,,
\end{equation}
where we have denoted the collection of those loops with $z_i=0$ as $T$ and the complementary one (with $x_j=0$) as $B$. 
\begin{figure}[H]
\centering
\includegraphics{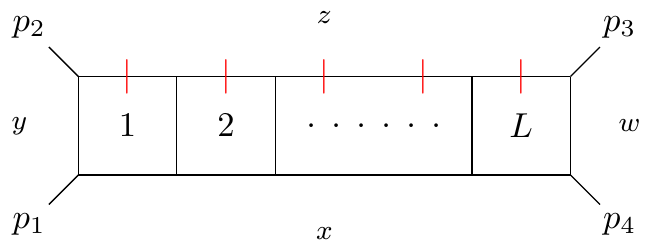}
\end{figure}
This trivial product is much simpler than the result for such cuts in $\mathcal{N}=4$ SYM case. For the latter, the ladder cut gets contribution from box ladder integrals and the next-to-ladder ones get contributions from many more topologies (such as tennis court at $L=3$). Here for the ABJM case, already the ladder cut gets contributions from ladder integrals with loop $1$ and loop $L$ being either box or triangle, and it remains an open question to understand such cuts (which trivially follow from geometry) from {\it e.g.} Feynman diagrams.  

One can consider more non-trivial cuts where {\it e.g.} $z_1=z_2=\cdots=z_{L{-}1}=w_L$ (similarly one can replace any number of $z$ by $x$ and/or $w$ by $y$). It is also very simple since only the star graph (with loop $L$ black and the rest white) contributes:
\begin{equation}
   \includegraphics[align=c,scale=1.2]{figures/star.pdf}\Rightarrow \includegraphics[align=c]{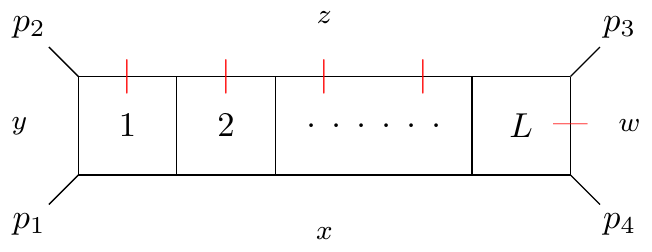}\,.
\end{equation}
From the closed form for star graph, \eqref{eq:starf}, the cut reads
\begin{equation}
    \frac{2^{L-1}}{y_L z_L\prod_{i=1}^{L-1}w_ix_i(-2+w_i y_L+x_i z_L)}\,.
\end{equation}
It would be very interesting to understand such cuts from physics, or use them as constraints on {\it e.g.} an ansatz for loop integrands. In the following, we will present two classes of cuts which we do understand from physics, unitarity cuts and vanishing cuts in ABJM theory. 
\subsection{Proof of perturbative unitarity for four-point amplitudes}\label{sec:uni}
The perturbative unitarity relates the discontinuity of the amplitude across a double cut to the product of two lower-loop ones. One can compute the discontinuity by taking the residue on the corresponding boundary of the amplituhedron and the optical theorem becomes a statement about the factorization of the residue on this boundary. Similar to the $\mathcal{N}=4$ case \cite{Arkani-Hamed:2013kca,YelleshpurSrikant:2019meu}, we now show that it emerges as a consequence of our four-point geometry. 

Let us begin with rewriting the optical theorem for four-point ABJM amplitude. We are interested in the case where one of the loops $A B$, cuts the lines $12$ and $34$ and all other loops (which we denote by $(A B)_i$ ) remain uncut. Thus we are calculating the residue of the $4$-point $L$-loop amplitude on the pole $\langle A B 12\rangle=\langle A B 34\rangle=0$. It is convenient to to parametrize the cut loop $AB$ as (\eqref{eq:parAB} with $y=w=0$) \begin{equation}\label{eq:cutAB}
    \begin{cases}
        & A=Z_1+x Z_2\\
        & B=Z_3+z Z_4\\
    \end{cases},
\end{equation}
and the symplectic condition $A\cdot\Omega\cdot B=0$ becomes $x z=1$ which gives $z=x^{-1}$; other uncut loops are still parametrized as \eqref{eq:parAB}.

If we compute the residue of the $L$-loop integrand $M_4^L(Z_1, Z_2, Z_3, Z_4)$ on this configuration, unitarity tells us that the result must be
\begin{equation}\label{eq:fac4}
\frac{d x}{x}\times \sum_{L_1+L_2=L-1} M_4^{L_1}\left(Z_1,A,B,Z_4\right) M_4^{L_2}\left(A,Z_2,Z_3,B\right).
\end{equation}
\begin{figure}[H]
    \centering
    \includegraphics[scale=2.5]{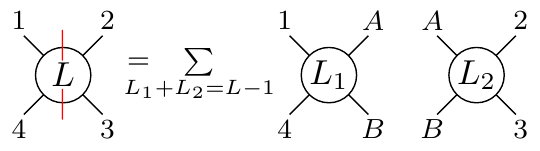}
\end{figure}

In~\cite{He:2022cup}, we have checked explicitly up to $L=5$ that the form of ABJM four-point amplituhedron satisfies \eqref{eq:fac4}. Now we propose a simple proof for all loops. From \eqref{eq:parAB} and \eqref{eq:cutAB}, the mutual positivity conditions between loops $AB$ and $(AB)_i$ becomes
\begin{equation}
    \langle AB (AB)_i\rangle=\frac{\langle 1234\rangle}{z}\left(z_i x^2-2x+x_i\right)<0,
\end{equation}
which implies that
\begin{equation}
    z_i x^2-2x+x_i<0\,,
\end{equation}
thus we either have
\begin{equation}\label{eq:L4x}
    x>\frac{1+\sqrt{w_ay_a}}{z_a}=\frac{x_a}{1-\sqrt{w_ay_a}},
\end{equation}
for some of the uncut loops $(AB)_a$, or
\begin{equation}\label{eq:R4x}
    0<x<\frac{1-\sqrt{w_\alpha y_\alpha}}{z_\alpha}=\frac{x_\alpha}{1+\sqrt{w_\alpha y_\alpha}},
\end{equation}
for the remaining uncut loops $(AB)_\alpha$.

If there are $L_1$ loops $(AB)_a$ satisfying \eqref{eq:L4x} and $L_2=L-L_1-1$ loops $(AB)_{\alpha}$ satisfying \eqref{eq:R4x}, we will show that $(AB)_a$ and $(AB)_\alpha$ satisfy all conditions for lower-loop ABJM amplituhedra as in \eqref{eq:fac4}.

For $L_1$ loops $(AB)_a$, we want to show that 
\begin{align}
    &\text{Tree Level: } \langle 1 A B 4\rangle=x\langle 1234\rangle<0\\\nonumber
&\text{Loop level: } \left\langle(A B)_a 1 A\right\rangle=x\left\langle(A B)_a 12\right\rangle<0, \left\langle(A B)_a A B\right\rangle<0, \left\langle(A B)_a B 4\right\rangle=\left\langle(A B)_a 34\right\rangle<0\\\nonumber
&\text{The sequence}\left\{\left\langle(A B)_a 1 A\right\rangle,\left\langle(A B)_a 1 B\right\rangle,\left\langle(A B)_a 14\right\rangle\right\} \text{ has } 2 \text{ sign flips}\\\nonumber
&\text{Mutual positivity: } \left\langle(A B)_a(A B)_b\right\rangle<0,
\end{align}
The only nontrivial one is the sign-flip condition, which means that $\langle (AB)_a14\rangle >0$. From \eqref{eq:L4x} we have $x>\frac{x_a}{1-\sqrt{y_a w_a}}>x_a$
\begin{equation}
    \langle (AB)_a14\rangle=\frac{x_a-x}{x}\langle1234\rangle>0.
\end{equation}

Similarly, for the $L_2$ loops $(AB)_{\alpha}$,we want to show that
\begin{align}
&\text{Tree Level: } \langle A 23 B\rangle=\frac{1}{x}\langle 1234\rangle<0\\\nonumber
&\text{Loop level: } \left\langle(A B)_\alpha A 2\right\rangle=\left\langle(A B)_\alpha 12\right\rangle<0, \left\langle(A B)_\alpha 23\right\rangle<0,\left\langle(A B)_\alpha 3 B\right\rangle=\left\langle(A B)_\alpha 34\right\rangle<0\\\nonumber
&\text{The sequence}\left\{\left\langle(A B)_\alpha A 2\right\rangle,\left\langle(A B)_\alpha A 3\right\rangle,\left\langle(A B)_\alpha A B\right\rangle\right\} \text{ has } 2 \text{ sign flips}\\\nonumber
&\text{Mutual positivity: } \left\langle(A B)_\alpha(A B)_\beta\right\rangle<0,
\end{align}
Similarly, the only non-trivial one to be shown is again the sign-flip condition,{\it{i.e.}} $\langle (AB)_{\alpha}A3\rangle>0$. From \eqref{eq:R4x}, we can see that 
\begin{equation}
    \langle (AB)_{\alpha}A3\rangle=(1-x z_\alpha)\langle 1234\rangle>0.
\end{equation}
To complete the proof of the factorization of ABJM four-point amplituhedron, we must show that the mutual positivity between the loops $(A B)_a$ and $(A B)_\alpha$ imposes no constraints. To see this, we can expand the loop $(A B)_a$ in terms of $\left\{Z_1, A, B, Z_4\right\}$ as
\begin{equation}
    \begin{cases}
        & A_a=Z_1+x_a A-w_a Z_4\\
        & B_a=y_a A+B+z_a Z_4,
    \end{cases}
\end{equation}

Then the mutual condition becomes
\begin{align}
\langle(AB)_a(AB)_\alpha\rangle&=\langle(AB)_\alpha23\rangle+x\langle(AB)_\alpha A2\rangle-2x\langle(AB)_\alpha2B\rangle\\\nonumber
&-(z^2+z)\langle(AB)_\alpha A3\rangle+z\langle(AB)_\alpha3B\rangle+z\langle (AB)_\alpha AB\rangle,
\end{align}
which is manifestly negative term by term.

In summary, we have shown that the four-point ABJM amplituhedron satisfies perturbative unitarity to all loops. The proof is basically the $D=3$ version of that for the amplituhedron in ${\cal N}=4$ SYM~\cite{Arkani-Hamed:2013kca,YelleshpurSrikant:2019meu} despite the fact that these geometries differ significantly (see next section). This shows the rigidity of the idea ``unitarity emerges from geometry" which could work for geometries in more general theories. 
\subsection{Bipartite graphs and vanishing cuts for ABJM}
Recall that in~\cite{He:2022cup}, we have shown that triple cuts and five-particle cuts, which contain vanishing odd-point amplitudes, vanish as a simple consequence of the geometry. This was already a strong indication that the geometry describes ABJM amplitudes. Now we want to show that any cut that vanishes geometrically, {\it i.e.} those that cannot be consistent with our bipartite geometry and its pole structure, must correspond to a cut which isolates an odd-point amplitude. Notice that all the disconnected graphs consist of bipartite connected components, so we only need to focus on the connected bipartite graph.

First we consider vanishing cuts with only mutual conditions $D_{i,j}=0$, which already illustrates how bipartite geometries fit perfectly with ABJM theory. The fact that geometrically any such vanishing cut must contain odd-point amplitudes (thus vanish in ABJM theory) is simply the following theorem~\cite{wiki:Plagiarism}:
\begin{theorem}
A graph is bipartite if and only if it does not contain any odd cycle.
\end{theorem}

For example, at $L=3$, we only have chain graphs which can be bipartite but no triangle graph since it cannot be bipartite. Similarly, we have an example for $L=4$ and $L=5$ (pentagon) that cannot be bipartite. 
\begin{figure}[H]
    \centering
    \includegraphics[scale=2]{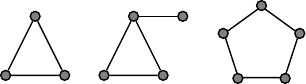}
\end{figure}

Since the ABJM four-point amplituhedron contains only bipartite graphs, any cut involving mutual conditions 
of the form $D_{i_1,i_2}=D_{i_2,i_3}=\cdots=D_{i_{2k},i_{2k+1}}=D_{i_{2k+1},i_1}=0$, (cutting an $(2k+1)$-gon) must vanish. Physically, we see that this cut must contain vanishing $(2k+1)$-point amplitude.

\begin{equation}
    \includegraphics[align=c,scale=1.5]{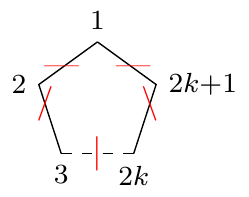}\Leftrightarrow\includegraphics[align=c,scale=1.5]{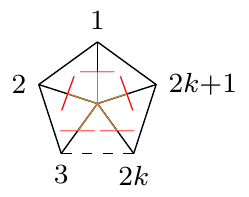}.
\end{equation}

Next we move to cuts that do not involve an odd cycle. For any such cut to vanish, it must contain at least two conditions of the form  
$\left(\ell_i\cdot j\right)=0$. show that our bipartite geometry, together with its pole structure, also guarantees that there's no odd particle cut physically.

Recall that the pole structure of bipartite graphs implies that we can not cut $x_i=0$ or $z_i=0$ for a black vertex $i$ and we can neither cut $w_j=0$ or $y_j=0$ for a white vertex $j$. 


In general, any vanishing cut including some external cuts must at least cut two external poles $(\ell_i\cdot j)$ and $(\ell_i^\prime\cdot j^\prime)$. If only one external pole $(\ell_i\cdot j)$ is cut, we can not determine whether vertex $i$ is black or white thus this cut does not vanish from geometry. And then since the graph is connected, there must exist at least one sub-chain connecting two vertexes $\ell_i,\ell_i^\prime$. The discussion can be divided into two cases according to whether the length of the sub-chain is even or odd.


\paragraph{Case I: }

If the length of the chain is even, say $2k$, and we cut the internal propagators $D_{1,2}=D_{2,3}=\cdots =D_{2k-1,2k}=0$. And we also cut, for example, $y_1=w_{2k}=0$ then geometrically, since the vertex $1$ and $2k$ are in a different color, so this cut must vanish. Physically, this cut corresponds to cutting the $2k$-chain into two $(2k+3)$-point amplitude $A_D(p_4,p_1,q_1,q_2q_3,\ldots,q_{2k},q)\times A_U(p_2,p_3,-q,-q_{2k},\ldots,-q_3,-q_2,-q_{1})$, so our bipartite pole structure once again ensures the vanishing property of ABJM odd point amplitude.

\begin{equation}
    \includegraphics[align=c,scale=1.2]{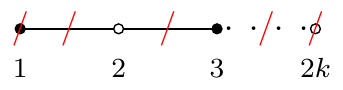}\Rightarrow\includegraphics[align=c,scale=1]{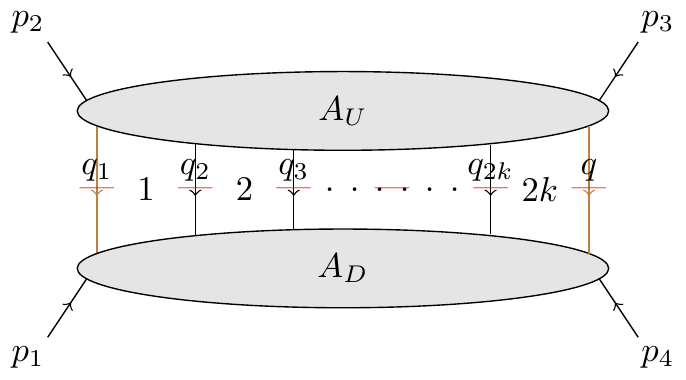}.
\end{equation}

\paragraph{Case II:}

If the length of the chain is odd, say $2k+1$, and we cut the internal propagators $D_{1,2}=D_{2,3}=\cdots =D_{2k,2k+1}=0$. And we also cut, for example, $y_1=x_{2k+1}=0$ then geometrically, since the vertex $1$ and $2k+1$ are in the same color, so this cut must vanish. Physically, this cut corresponds to cutting the $2k+1$-chain into a $(2k+3)$-point amplitude and a $(2k+5)$-point amplitude $A_D(p_1,q_1,q_2,\ldots,q_{2k-1},q)\times A_U(p_2,-q_1,-q_2,\ldots,-q_{2k{-}1},p_3,p_4,-q)$, again corresponding to vanishing odd-point ABJM amplitudes. 

\begin{equation}
    \includegraphics[align=c,scale=1.2]{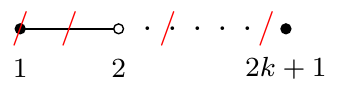}\Rightarrow\includegraphics[align=c,scale=1]{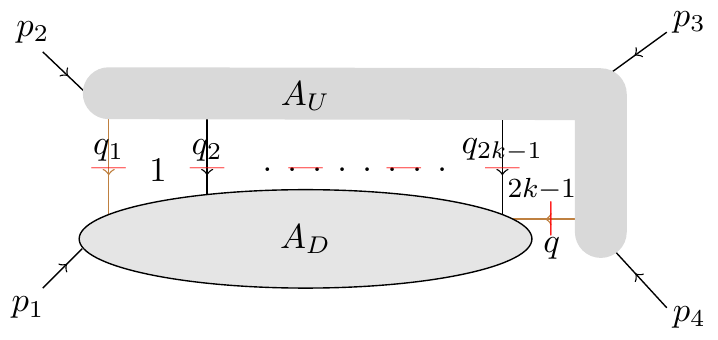}.
\end{equation}

Therefore, our bipartite geometry along with its pole structure nicely ensures that any cut which isolates an odd-point amplitude must vanish. 

Before proceeding, we remark that these bipartite graphs clearly appear for negative geometries associated with ABJM amplituhedron to all multiplicities. The reason is that any higher-point integrand is contained in cuts of (higher-loop) four-point ones, and as a consequence, the mutual negative conditions must still form bipartite graphs, independent of other conditions. Therefore, we expect these bipartite negative geometries, which are very special to ABJM theory, to contain complete information about ``mutual conditions" for ABJM integrands with any number of loops and legs.

\section{Integrating negative geometries for ABJM amplitudes}\label{sec:integration}
In this section, we consider integrating the forms of negative geometries for four-point ABJM amplituhedron. We emphasize that for each (connected) negative geometry, one can integrate all but one loop variables, which result in an infrared-finite function; if we combine them together and perform the last loop integration, we obtain the (divergent) logarithm of the amplitude which contains information about $\Gamma_{\rm cusp}$. Just like in ${\cal N}=4$ SYM, it is already interesting to consider integrating the forms of individual negative geometry. Before proceeding, let us recall the $L=1$ case where we do no integration:

\begin{eqnarray}
\vcenter{\hbox{\scalebox{0.9}{
\tikzset{every picture/.style={line width=0.75pt}} 

\begin{tikzpicture}[x=0.75pt,y=0.75pt,yscale=-1,xscale=1]

\draw  [color={rgb, 255:red, 0; green, 0; blue, 0 }  ,draw opacity=1 ][fill={rgb, 255:red, 0; green, 0; blue, 0 }  ,fill opacity=1 ][line width=0.75]  (135.5,136.7) -- (143.5,136.7) -- (143.5,144.7) -- (135.5,144.7) -- cycle ;


\end{tikzpicture}

}
}}
        &&=\frac{\epsilon(\ell_1,1,2,3,4)}{(\ell_1\cdot 1)(\ell_1\cdot 2)(\ell_1\cdot 3)(\ell_1 \cdot 4)}.
\end{eqnarray}
Here, we use the box to represent the loop is unintegrated.

\subsection{Integrated results up to $L=3$}
Since our integrand contains even/odd number of $\epsilon$ in the numerator at even/odd loop, we expect that the integrated result (with one loop frozen) should be proportional to such an $\epsilon$ factor for odd $L$, and contains no such factor for even $L$. Also taking into account DCI weight and cyclic symmetry of the amplitude, we propose to use the following normalization for the finite-function ${\cal W}_L$ at odd and even $L$ respectively:
\begin{equation}\label{FL-1}
{\cal W}_L=\begin{cases}
\displaystyle \frac{\epsilon (\ell_1,1,2,3,4)}{(\ell_1 \cdot 1)(\ell_1 \cdot 2) (\ell_1 \cdot 3) (\ell_1 \cdot 4)}~F_{L-1}(z)\,, \quad L~{\rm odd}\\
\displaystyle \left(\frac{(1\cdot 3) (2\cdot 4)}{(\ell_1\cdot 1) (\ell_1\cdot 2) (\ell_1\cdot 3) (\ell_1\cdot 4)} \right)^{3/4}~F_{L-1}(z), \quad L~{\rm even}
\end{cases}
\end{equation}
where we have denoted the unintegrated loop variable as $\ell_1$, and the normalized function, $F_{L-1}$ only depends on the cross-ratio
\begin{equation}
    z=\frac{(\ell_1\cdot 2) (\ell_1\cdot 4) (1\cdot 3)}{(\ell_1\cdot 1) (\ell_1\cdot 3) (2\cdot 4)}\,.
\end{equation}

In this normalization, we see $F_0(z)=1$. At this stage we do not know what kind of functions $F_{L-1} (z)$ are, except that it must be cyclically invariant, so $F(z)=F(1/z)$ (since $i\to i{+}1$ amounts to $z \to 1/z$). We will see after integration that $F_{L-1}$ turn out to be uniform transcendental functions of weight $L-1$ at least for $L\leq 3$.

To determine $F_{L-1}(z)$, we first review how to integrate a massive one-loop triangle integral which is the most basic operation we have on these integrals:
\begin{equation}
    \int_\ell \frac{1}{(\ell\cdot i)(\ell\cdot j)(\ell\cdot k)}=2 \int_0^\infty \frac{[d^2a_i a_j a_k]}{\text{vol}(\text{GL}(1))}\int_\ell \frac{1}{(\ell\cdot A)^3},
\end{equation}
where $A=a_i X_i+a_j X_j+a_k X_k$ with three Feynman parameters $a_i$, $a_j$ and $a_k$, and the measure $\frac{[d^{n-1}\alpha_1 \alpha_2\dots \alpha_n]}{\operatorname{vol}(G L(1))}\equiv d \alpha_0 \cdots d \alpha_n \delta\left(\alpha_i-1\right)$ (for any $\left.i\right)$. After integrating $D=3$ loop momentum\footnote{In the calculation that follows, we have adopted the convention of~\cite{Caron-Huot:2012sos} and omitted a factor of $1/(4\pi)$ for each loop in order to avoid clutter.}, 
\begin{equation}
    \int_\ell \frac{1}{(\ell\cdot A)^3}=\frac{1}{8(\frac{1}{2}(A\cdot A))^{3/2}}=\frac{1}{8\big(a_i a_j (i\cdot j){+}a_i a_k (i\cdot k){+}a_j a_k (j\cdot k)\big)^{3/2}},
\end{equation}
we obtain the well-known result~\cite{Caron-Huot:2012sos, Brandhuber:2012un,Brandhuber:2012wy}:
\begin{equation}\label{triangle}
    \begin{split}
        \int_\ell \frac{1}{(\ell\cdot i)(\ell\cdot j)(\ell\cdot k)}
        &= \frac{1}{4} \int_0^\infty \frac{[d^2a_i a_j a_k]}{\text{vol}(\text{GL}(1))} \frac{1}{\big(a_i a_j (i\cdot j){+}a_i a_k (i\cdot k){+}a_j a_k (j\cdot k)\big)^{3/2}}\\
        &=\frac{1}{2}\frac{\pi}{\sqrt{(i\cdot j)}\sqrt{(j\cdot k)}\sqrt{(k\cdot i)}}.
    \end{split}
\end{equation}

\paragraph{Two loop.} For $L=2$, we have either black or white node being $\ell_1$ (frozen), and it is trivial to integrate out $\ell_2$ since it corresponds to a one-loop triangle integral: 
\begin{eqnarray}
\vcenter{\hbox{\scalebox{0.9}{
\begin{tikzpicture}[x=0.75pt,y=0.75pt,yscale=-1,xscale=1]

\draw  [color={rgb, 255:red, 0; green, 0; blue, 0 }  ,draw opacity=1 ][fill={rgb, 255:red, 0; green, 0; blue, 0 }  ,fill opacity=1 ][line width=0.75]  (135.5,136.7) -- (143.5,136.7) -- (143.5,144.7) -- (135.5,144.7) -- cycle ;
\draw [line width=0.75]    (139.5,140.7) -- (219.8,140.7) ;
\draw  [fill={rgb, 255:red, 255; green, 255; blue, 255 }  ,fill opacity=1 ] (215.72,141.06) .. controls (215.52,138.8) and (217.19,136.81) .. (219.44,136.62) .. controls (221.7,136.42) and (223.69,138.09) .. (223.88,140.34) .. controls (224.08,142.6) and (222.41,144.59) .. (220.16,144.78) .. controls (217.9,144.98) and (215.91,143.31) .. (215.72,141.06) -- cycle ;

\draw (132.5,151.33) node [anchor=north west][inner sep=0.75pt]  [font=\footnotesize] [align=left] {$\displaystyle 1$};
\draw (213,151.33) node [anchor=north west][inner sep=0.75pt]  [font=\footnotesize] [align=left] {$\displaystyle 2$};

\end{tikzpicture}

}
}}
        &&=-2\frac{(1\cdot 3)(2\cdot 4)}{(\ell_1\cdot 2)(\ell_1\cdot 4)}\int_{\ell_2} \frac{1}{(\ell_2 \cdot \ell_1) (\ell_2\cdot 1)(\ell_2 \cdot 3)}\nonumber\\
        &&=-\frac{\sqrt{(1\cdot 3) }(2\cdot 4)}{\sqrt{(\ell_1\cdot 1)}\sqrt{(\ell_1\cdot 3)}(\ell_1\cdot 2)(\ell_1\cdot 4)}\times \pi\nonumber\\
        &&=-\left(\frac{(1\cdot 3)(2\cdot 4)}{(\ell_1\cdot 1)(\ell_1\cdot 2)(\ell_1\cdot 3)(\ell_4\cdot 4)}\right)^{\frac{3}{4}}\times \pi z^{1/4},
\end{eqnarray}
and similarly for the second graph with white and black nodes swapped with $z\leftrightarrow z^{-1}$. In the last step we have pulled out the prefactor as defined in \eqref{FL-1}, and the final result is 
\begin{equation}
F_1(z)=-\pi (z^{1/4} + z^{-1/4}).
\end{equation}

\paragraph{Three loop.} For $L=3$, we have two inequivalent cases: the unintegrated loop in the middle of the chain or at one end. For the former the integration for the other two loops is again trivial (two triangle integrals) and we obtain
\begin{eqnarray}
\vcenter{\hbox{\scalebox{0.9}{
\begin{tikzpicture}[x=0.75pt,y=0.75pt,yscale=-1,xscale=1]

\draw [line width=0.75]    (139.5,140.7) -- (219.8,140.7) ;
\draw  [fill={rgb, 255:red, 0; green, 0; blue, 0 }  ,fill opacity=1 ] (215.72,141.06) .. controls (215.52,138.8) and (217.19,136.81) .. (219.44,136.62) .. controls (221.7,136.42) and (223.69,138.09) .. (223.88,140.34) .. controls (224.08,142.6) and (222.41,144.59) .. (220.16,144.78) .. controls (217.9,144.98) and (215.91,143.31) .. (215.72,141.06) -- cycle ;
\draw  [fill={rgb, 255:red, 0; green, 0; blue, 0 }  ,fill opacity=1 ] (135.4,140.7) .. controls (135.4,138.44) and (137.24,136.6) .. (139.5,136.6) .. controls (141.76,136.6) and (143.6,138.44) .. (143.6,140.7) .. controls (143.6,142.96) and (141.76,144.8) .. (139.5,144.8) .. controls (137.24,144.8) and (135.4,142.96) .. (135.4,140.7) -- cycle ;
\draw  [color={rgb, 255:red, 0; green, 0; blue, 0 }  ,draw opacity=1 ][fill={rgb, 255:red, 255; green, 255; blue, 255 }  ,fill opacity=1 ][line width=0.75]  (176.5,136.7) -- (184.5,136.7) -- (184.5,144.7) -- (176.5,144.7) -- cycle ;

\draw (132.5,152.33) node [anchor=north west][inner sep=0.75pt]  [font=\footnotesize] [align=left] {$\displaystyle 2$};
\draw (213,152.33) node [anchor=north west][inner sep=0.75pt]  [font=\footnotesize] [align=left] {$\displaystyle 3$};
\draw (175,152.33) node [anchor=north west][inner sep=0.75pt]  [font=\footnotesize] [align=left] {$\displaystyle 1$};

\end{tikzpicture}
}
}}
        &&=\frac{4\, \epsilon(\ell_1,1,2,3,4)(2\cdot 4)}{(\ell_1\cdot 1)(\ell_1\cdot 3)}\int_{\ell_2,\ell_3} \frac{1}{(\ell_2\cdot \ell_1)(\ell_2\cdot 2)(\ell_2\cdot 4)(\ell_3\cdot \ell_1)(\ell_3 \cdot 2)(\ell_3\cdot 4)}\nonumber\\
        &&=\frac{\epsilon(\ell_1,1,2,3,4)}{(\ell_1\cdot 1)(\ell_1\cdot 2)(\ell_1\cdot 3)(\ell_1\cdot 4)}\times \pi^2. 
\end{eqnarray}

For the latter, it is the only non-trivial computation we need up to $L=3$. Let us present the details of the calculation.
\begin{eqnarray}\label{eq:integration_2-link}
\vcenter{\hbox{\scalebox{0.9}{
\begin{tikzpicture}[x=0.75pt,y=0.75pt,yscale=-1,xscale=1]

\draw  [color={rgb, 255:red, 0; green, 0; blue, 0 }  ,draw opacity=1 ][fill={rgb, 255:red, 0; green, 0; blue, 0 }  ,fill opacity=1 ][line width=0.75]  (135.5,136.7) -- (143.5,136.7) -- (143.5,144.7) -- (135.5,144.7) -- cycle ;
\draw [line width=0.75]    (139.5,140.7) -- (219.8,140.7) ;
\draw  [fill={rgb, 255:red, 255; green, 255; blue, 255 }  ,fill opacity=1 ] (176.42,141.06) .. controls (176.22,138.8) and (177.89,136.81) .. (180.14,136.62) .. controls (182.4,136.42) and (184.39,138.09) .. (184.58,140.34) .. controls (184.78,142.6) and (183.11,144.59) .. (180.86,144.78) .. controls (178.6,144.98) and (176.61,143.31) .. (176.42,141.06) -- cycle ;
\draw  [fill={rgb, 255:red, 0; green, 0; blue, 0 }  ,fill opacity=1 ] (215.72,141.06) .. controls (215.52,138.8) and (217.19,136.81) .. (219.44,136.62) .. controls (221.7,136.42) and (223.69,138.09) .. (223.88,140.34) .. controls (224.08,142.6) and (222.41,144.59) .. (220.16,144.78) .. controls (217.9,144.98) and (215.91,143.31) .. (215.72,141.06) -- cycle ;

\draw (134,151.33) node [anchor=north west][inner sep=0.75pt]  [font=\footnotesize] [align=left] {$\displaystyle 1$};
\draw (214,151.33) node [anchor=north west][inner sep=0.75pt]  [font=\footnotesize] [align=left] {$\displaystyle 3$};
\draw (174.5,151.33) node [anchor=north west][inner sep=0.75pt]  [font=\footnotesize] [align=left] {$\displaystyle 2$};

\end{tikzpicture}
}
}}
        &&=\frac{4\,(2\cdot 4)}{(\ell_1\cdot 2)(\ell_1\cdot 4)}\int_{\ell_2,\ell_3} \frac{\epsilon(\ell_2,1,2,3,4)}{(\ell_2\cdot \ell_1)(\ell_2\cdot 1)(\ell_2\cdot 3)(\ell_2\cdot \ell_3)(\ell_3 \cdot 2)(\ell_3\cdot 4)}.
\end{eqnarray}
As a standard procedure for performing the Feynman integral, we begin with introducing the Feynman parameters. A nice way to introduce the Feynman parameters for the two-loop integrals can follow the prescription of~\cite{Caron-Huot:2012awx,Caron-Huot:2012sos}. Let us omit the prefactor that does not affect integration and focus on the non-trivial integral for now. The integral~\eqref{eq:integration_2-link} in  Feynman parameters space is
\begin{equation}\label{eq:two-link_feynman}
    \begin{split}
        &\int_{\ell_2,\ell_3} \frac{\epsilon(\ell_2,1,2,3,4)}{(\ell_2\cdot \ell_1)(\ell_2\cdot 1)(\ell_2\cdot 3)(\ell_2\cdot \ell_3)(\ell_3 \cdot 2)(\ell_3\cdot 4)}\\
        =& 2 \int_0^\infty \frac{[d^2  a_1 a_3 a_{\ell_1}]}{\text{vol}\left(\text{GL}(1)\right)} \frac{[d^2 b_2 b_4]}{\text{vol}\left(\text{GL}(1)\right)} \int_{\ell_2,\ell_3} \frac{\epsilon(\ell_2,1,2,3,4)}{(\ell_2 \cdot A)^3 (\ell_2\cdot \ell_3)(\ell_3 \cdot B)^2}\\
        =& - \int_0^\infty \frac{[d^2  a_1 a_3 a_{\ell_1}]}{\text{vol}\left(\text{GL}(1)\right)} \frac{[d^2 b_2 b_4]}{\text{vol}\left(\text{GL}(1)\right)} \int_{\ell_2,\ell_3} \frac{\epsilon(\partial_A,1,2,3,4)}{(\ell_2 \cdot A)^2 (\ell_2\cdot \ell_3)(\ell_3 \cdot B)^2}\\
        =& -\int_0^\infty \frac{dc}{4\pi \sqrt{c}} \int_0^\infty \frac{[d^5  a_1 a_3 a_{\ell_1} b_2 b_4]}{\text{vol}(\text{GL}(1))} \frac{\epsilon(\partial_A,1,2,3,4)}{\left((c+1)\frac{1}{2}A^2 +A\cdot B+\frac{1}{2} B^2\right)^2},
    \end{split}
\end{equation}
where $A=a_{\ell_1} X_{\ell_1}+a_1 X_1+a_3 X_3$, $B= b_2 X_2+b_4 X_4$. The Feynman parameters $a_i$ and $b_i$ are associated with the propagators of the integral, whereas $c$ is not related to any of them. Although we could eliminate the newly introduced Feynman parameter $c$ right away, leaving it in the equation will enable us to delay addressing square roots until the final step.

After evaluating inner products and doing a simple rescaling of the integration variables by
\begin{equation}
 \begin{gathered}
        a_1^\prime=\frac{(1\cdot 3)(\ell_1\cdot 1)}{(\ell_1\cdot 3)} a_1,\quad a_3^\prime =(1\cdot 3) a_3, \quad a_{\ell_1}^\prime=(\ell_1\cdot a)a_{\ell_1}^\prime,\\
        b_2^\prime=\frac{(2\cdot 4)(\ell_1\cdot 1)}{(\ell_1 \cdot 4)}b_2, \quad b_4^\prime=\frac{(2\cdot 4)(\ell_1\cdot 1)}{(\ell_1 \cdot 2)}b_4,
    \end{gathered}
\end{equation}
the integral~\eqref{eq:two-link_feynman} takes a very compact form: after stripping off the factor $\epsilon(a,1,2,3,4)/(1\cdot 3)(\ell_1\cdot 2)(\ell_1\cdot 4)$ (we will put it back after obtaining the integrated result of~\eqref{eq:integration_2-link}), it becomes a function of the cross-ratio $z$ only:
\begin{equation}
    \begin{split}
       \int_0^\infty \frac{dc}{4 \pi \sqrt{c}}\int_0^\infty \frac{[d^5  a_1^\prime a_3^\prime a_{\ell_1}^\prime b_2^\prime b_4^\prime]}{\text{vol}(\text{GL}(1))}   \frac{2 a_{\ell_1}^\prime(1{+}c) /z}{ \left( b_2^\prime b_4^\prime{+}a_1^\prime a_3^\prime (1{+}c)/z +a_{\ell_1}^\prime(b_2^\prime{+}b_4^\prime{+}(a_1^\prime{+}a_3^\prime)(1+c)/z)\right)^3}.
    \end{split}
\end{equation}
To evaluate it, we simply integrate over the variable $a_i^\prime$, $b_i^\prime$ one at a time. During each integration step, the integral can be expressed in terms of rational factors of the form $dx/(x-x_i)^n$ with $n\ge 1$, multiplied by logarithms or polylogarithms whose arguments are ratios of functions at most linear in $x$. Such integrals can be performed recursively, and the result can be expressed in term of polylogarithm~\cite{Caron-Huot:2011dec} (we follow the algorithm proposed in \cite{Caron-Huot:2012awx} and the code in \cite{He:2020uxy,Gint}). After integrating out all the Feynman parameters $a_i^\prime$, $b_i^\prime$, we arrive at the $c$-integrand
\begin{equation}
    \begin{split}
        \frac{4\,\epsilon(\ell_1,1,2,3,4)}{(\ell_1\cdot 1)(\ell_1\cdot 2)(\ell_1\cdot 3)(\ell_1\cdot 4)}\int_0^\infty \frac{dc}{4 \pi \sqrt{c}}\frac{1}{1+(1+c)/z}\left(\frac{\pi^2}{2}+\frac{1}{2}\log^2{\frac{1+c}{z}}\right).
    \end{split}
\end{equation}
Here, we have put back the prefactors $\frac{4(2\cdot 4)}{(\ell_1\cdot 2)(\ell_1\cdot 4)}$ and $\frac{\epsilon(\ell_1,1,2,3,4)}{(1\cdot 3)(\ell_1\cdot 2)(\ell_1\cdot 4)}$. 

The final $c$ integral appears distinct from the integrals encountered in the previous steps, but after changing the variable $d=\sqrt{c}$ the integral is still the form dlog multiplied by logarithms or polylogarithms. The full result in \eqref{eq:integration_2-link} is again of the form of \eqref{FL-1} and we have 
\begin{equation}
\frac{2\, \epsilon(\ell_1,1,2,3,4)}{(\ell_1\cdot 1)(\ell_1\cdot 2)(\ell_1\cdot 3)(\ell_1\cdot 4)}\times  f\left(\frac{1}{z}\right),
\end{equation}
where we have defined for $x>0$
\begin{equation}\label{fx}
f(z):=\frac{t-1}{t+1}\biggl(\frac{\pi ^2}{2}+ \operatorname{Li}_2(1-t)+\log (t)\log (t{-}1)- \frac{1}{4} \log (t)^2 \biggr)
\,\, \text{with $t:=\frac{\sqrt{1+z}+\sqrt{z}}{\sqrt{1+z}-\sqrt{z}}$}.
\end{equation}
where note the prefactor is simply $\sqrt{\frac{x}{1+x}}$. Taking into account the fact that we have $6$ graphs in total, the final result for $F_2$ (see \eqref{FL-1}) is 
\begin{equation}
F_2(z)=4\left( f(z) + f\biggl(\frac 1 z\biggr)+ \frac{\pi^2}{2}\right).
\end{equation}
Note that the weight-$2$ function $F_2(z)$ is not pure: apart from the constant $2\pi^2$, $f(z)$ and $f(\frac 1 z)$ have different prefactors; the two prefactors are 
\begin{equation}
\sqrt{\frac{z}{1+z}}=\frac{2 \lambda}{1+\lambda^2} \quad {\rm and} \quad \sqrt{\frac 1{1+z}}=\frac{1-\lambda^2}{1+\lambda^2}\,,
\end{equation}
with the change of variable $\sqrt{z}\equiv \frac{2 \lambda}{1-\lambda^2}$ for $0<\lambda<1$. The symbols~\cite{Goncharov:2010jf,Duhr:2011zq,Duhr:2012fh} of the pure functions accompanying these two prefactors in $f(z)$ and $f(\frac 1 z)$ are
\begin{equation}
\frac{\lambda}{1-\lambda^2} \otimes \frac{(1+\lambda)^2}{(1-\lambda)^2} \quad {\rm and} \quad \frac{\lambda}{1-\lambda^2} \otimes \lambda^2 \,,
\end{equation}
respectively. The alphabet of $F_2(z)$ consists of $\lambda, 1-\lambda$ and $1+\lambda$. 

\subsection{Cusp anomalous dimension from integrated results}

Let us extract the cusp anomalous dimension from the functions obtained from the integration of  negative geometry. This can be done by integrating out the final loop variable: $\mathcal{W}_{L}$ diverges and the cusp anomalous dimension is encoded in the coefficient of $1/\epsilon^2$ when the integral is done in $D=3- 2 \epsilon$. We follow the steps in~\cite{Arkani-Hamed:2021iya} to obtain $\Gamma_{\rm cusp}$.

To evaluate the last loop integration and extract $\epsilon^{-2}$ divergence, one can expand $F_{L-1}(z)$ in $z$ around $z=0$. This series expansion has logarithmic divergence in general: 
\begin{equation}
    F_{L-1}(z)=\sum_{p,q}c^{(L-1)}_{p,q}z^p\log(z)^q,
\end{equation}
where $q$ is a non-negative integer, and $p$ can be any rational number. However, we only need to calculate the integral for $z^p$ since 
$z^p\log(z)^q=\frac{\partial^q}{\partial p^q}z^p$. 

First we consider the case with $L$ even:
\begin{equation}
    \begin{split}
        \int_{\ell_1} \left(\frac{(1\cdot 3)(2\cdot 4)}{(\ell_1\cdot 1)(\ell_1\cdot 2)(\ell_1\cdot 3)(\ell_1\cdot 4)}\right)^{\frac{3}{4}} z^p
        =\, & (1\cdot 3)^{\frac{3}{4}+p}(2\cdot 4)^{\frac{3}{4}-p}\frac{\Gamma(3) }{\Gamma\left(\frac{3}{4}+p\right)^2 \Gamma\left(\frac{3}{4}-p\right)^2}  \\
        & \,\,\times \int \frac{[d^3 a_1 a_2 a_3 a_4]}{\text{vol}\left(\text{GL}(1)\right)} \int_{\ell_1}\frac{(a_1 a_3)^{-\frac{1}{4}+p} (a_2 a_4)^{-\frac{1}{4}-p}}{(\ell_1\cdot A)^3} . 
    \end{split}
\end{equation}
After integrating out the loop $\ell_1$ and rescaling the Feynman parameters by $a_3$, $a_4$ by $1/(1\cdot 3)$, $1/(2\cdot 4)$ respectively, the integral becomes     
\begin{equation}\label{eq:cusp_even_step}
\begin{split}
    & \frac{\Gamma(3)}{\Gamma\left(\frac{3}{4}+p\right)^2 \Gamma\left(\frac{3}{4}-p\right)^2} \int \frac{[d^3 a_1 a_2 a_3 a_4]}{\text{vol}\left(\text{GL}(1)\right)} \frac{(a_1 a_3)^{-\frac{1}{4}+p} (a_2 a_4)^{-\frac{1}{4}-p}}{8 \big(a_1 a_3 +a_2 a_4 \big)^{3/2}}.
\end{split} 
\end{equation}

Now we go to $D=3-2\epsilon$, and by exploiting GL(1) invariance the coefficient of $\log^2$ divergence is given by setting $a_i=1$ for $i=2,3,4$:
\begin{equation}
         \frac{\Gamma(3)}{\Gamma\left(\frac{3}{4}+p\right)^2 \Gamma\left(\frac{3}{4}-p\right)^2} \int_0^\infty  \frac{da_1 a_1^{-1/4+p}}{8 \big(1+a_1  \big)^{3/2-\epsilon}}
        = \frac{1}{2\sqrt{\pi}} \frac{1}{\Gamma\left(\frac{3}{4}-p\right)\Gamma\left(\frac{3}{4}+p\right)}.
\end{equation}
Therefore, the cusp anomalous dimension at $L=2$ (with factor $(N/k)^2$ suppressed) is
\begin{equation}
\begin{split}
    \mathcal{I}_e\left(F_1(z)\right)&=-\pi\left( \mathcal{I}_e(z^{1/4}) + \mathcal{I}_e (z^{-1/4})\right)=-1.
\end{split}
\end{equation}
where the operation $\mathcal{I}_e$ is to extract the coefficient of $1/\epsilon^2$ when $L$ is even. This agrees with the coefficient of $1/\epsilon^2$ found in the calculation of the two-loop amplitudes~\cite{Chen:2011vv,Caron-Huot:2012sos}. 

We can use the same method to extract the cusp anomalous dimension for odd $L$
\begin{equation}\label{eq:odd_integration}
    \begin{split}
        \int_{\ell_1} \frac{\epsilon(\ell_1,1,2,3,4)}{(\ell_1\cdot 1)(\ell_1\cdot 2)(\ell_1\cdot 3)(\ell_1\cdot 4)} z^p
        =& -\frac{(1\cdot 3)^p}{(2\cdot 4)^p} \frac{\Gamma(3)}{\Gamma(1+p)^2\Gamma(1-p)^2} \int \frac{[d^3 a_1 a_2 a_3 a_4]}{\text{vol}(\text{GL}(1))}\\
        & \times \int_{\ell_1} \epsilon(\partial_A,1,2,3,4)\frac{1}{(\ell_1\cdot A)^3} \left(\frac{a_1 a_3}{a_2 a_4}\right)^p,
    \end{split}
\end{equation}
where $A=\sum_{i} a_i X_i$. Let us first focus on the inner integral and only consider the relevant factor for the integration $\ell_1$. For the purpose of extracting the terms survive from $\epsilon(\partial_A,1,2,3,4)$, we go to  the $D=3-2\epsilon$ dimension and introduce infinity point $X_I=(\vec{0}_D,0,1)$. After integrating $\ell_1$, we obtain
\begin{equation}
    \begin{split}
        &\epsilon(\partial_A,1,2,3,4) \int \frac{d^{D+2} \ell_1 \delta(\ell_1^2)}{\text{vol}(\text{GL}(1))} \frac{1}{(\ell_1\cdot A)^3(a\cdot I)^{D-3}}=\frac{\epsilon(\partial_A,1,2,3,4)}{8(A\cdot I)^{D-3}(\frac{1}{2}A^2)^{3-\frac{D}{2}}}\\
        =& \left(D/2-4\right)\frac{\epsilon(A,1,2,3,4)}{8(A\cdot I)^{D-3}(\frac{1}{2}A^2)^{4-\frac{D}{2}}}-(D-3) \frac{\epsilon(I,1,2,3,4)}{8(A\cdot I)^{D-2} ( \frac{1}{2}A^2)^{3-\frac{D}{2}}}.
    \end{split}
\end{equation}
The first term vanishes because $A$ is a linear combination of four external legs. We plug the second term into~\eqref{eq:odd_integration} and do the same resale of $a_3$ and $a_4$ in~\eqref{eq:cusp_even_step}. After remove GL(1) and two additional GL(1) invariances ($a_2=a_3=a_4=1$),  the integral gives an order $\epsilon$ contribution to the cusp constant:
\begin{equation}
\begin{split}
    &  \frac{2\, \epsilon\, \Gamma(3)\, \epsilon(I,1,2,3,4)}{\Gamma\left(\frac{3}{4}+p\right)^2 \Gamma\left(\frac{3}{4}-p\right)^2} \int_0^\infty  \frac{da_1 a_1^{-1/4+p} }{8 \big(1 +a_1 \big)^{3/2-\epsilon}} = \frac{\epsilon\, \Gamma\left(\frac{3}{4}-p-\epsilon\right)\,\epsilon(I,1,2,3,4)}{4\Gamma\left(\frac{3}{4}-p\right)^2\Gamma\left(\frac{3}{4}+p\right)\Gamma\left(\frac{3}{2}-\epsilon\right)}\sim O(\epsilon).
\end{split} 
\end{equation}
Therefore, the cusp anomalous dimension at the odd $L$ is always zero:
\begin{eqnarray}
    \mathcal{I}_o (F_{L-1}(z))=0.
\end{eqnarray}
Again, the operator $\mathcal{I}_o$ is to extract the $1/\epsilon^2$ term when $L$ is odd.

\subsection{A reduction identity for integrating negative geometries}

As mentioned before, a special property for ABJM negative geometries, is that the simplest integration one can trivially do is to integrate out a massive triangle, for any ``leaf" (valency-$1$ node) of the bipartite graph. This provides a simple reduction identity which relates higher-loop negative geometries to lower-loop ones by integrating all ``leaves". For example, one can trivially integrate out all $L{-}1$ leaves for any $L$-loop star graph: 
\begin{eqnarray}
\vcenter{\hbox{\scalebox{1.0}{
\begin{tikzpicture}[x=0.75pt,y=0.75pt,yscale=-1,xscale=1]

\draw [line width=0.75]    (146.63,168.33) -- (177.14,198.89) ;
\draw [line width=0.75]    (146.63,168.33) -- (177.14,137.47) ;
\draw [line width=0.75]    (146.63,168.33) -- (188.49,156.86) ;
\draw  [fill={rgb, 255:red, 0; green, 0; blue, 0 }  ,fill opacity=1 ] (173.92,137.47) .. controls (173.92,135.68) and (175.36,134.24) .. (177.14,134.24) .. controls (178.93,134.24) and (180.37,135.68) .. (180.37,137.47) .. controls (180.37,139.25) and (178.93,140.7) .. (177.14,140.7) .. controls (175.36,140.7) and (173.92,139.25) .. (173.92,137.47) -- cycle ;
\draw  [fill={rgb, 255:red, 0; green, 0; blue, 0 }  ,fill opacity=1 ] (185.26,156.86) .. controls (185.26,155.08) and (186.71,153.63) .. (188.49,153.63) .. controls (190.27,153.63) and (191.72,155.08) .. (191.72,156.86) .. controls (191.72,158.64) and (190.27,160.09) .. (188.49,160.09) .. controls (186.71,160.09) and (185.26,158.64) .. (185.26,156.86) -- cycle ;
\draw  [fill={rgb, 255:red, 0; green, 0; blue, 0 }  ,fill opacity=1 ] (173.92,198.89) .. controls (173.92,197.1) and (175.36,195.66) .. (177.14,195.66) .. controls (178.93,195.66) and (180.37,197.1) .. (180.37,198.89) .. controls (180.37,200.67) and (178.93,202.12) .. (177.14,202.12) .. controls (175.36,202.12) and (173.92,200.67) .. (173.92,198.89) -- cycle ;
\draw  [fill={rgb, 255:red, 255; green, 255; blue, 255 }  ,fill opacity=1 ] (143.4,168.33) .. controls (143.4,166.55) and (144.85,165.11) .. (146.63,165.11) .. controls (148.41,165.11) and (149.86,166.55) .. (149.86,168.33) .. controls (149.86,170.12) and (148.41,171.56) .. (146.63,171.56) .. controls (144.85,171.56) and (143.4,170.12) .. (143.4,168.33) -- cycle ;


\draw (86,151) node [anchor=north west][inner sep=0.75pt]   [align=left] {$\displaystyle \int _{1,\cdots ,L}$};



\draw (133.58,162.22) node [anchor=north west][inner sep=0.75pt]  [font=\footnotesize] [align=left] {$\displaystyle 1$};
\draw (170,161) node [anchor=north west][inner sep=0.75pt]   [align=left] {$\displaystyle \vdots $};
\draw (183.04,123.15) node [anchor=north west][inner sep=0.75pt]  [font=\footnotesize] [align=left] {$\displaystyle 2$};
\draw (194.32,149.4) node [anchor=north west][inner sep=0.75pt]  [font=\footnotesize] [align=left] {$\displaystyle 3$};
\draw (180.52,199.67) node [anchor=north west][inner sep=0.75pt]  [font=\footnotesize] [align=left] {$\displaystyle L$};



\draw (210,145) node [anchor=north west][inner sep=0.75pt]   [align=left] {$\displaystyle =\int _{1} \frac{1}{t_1}\,
 \begin{cases}
 s_1^{\frac{L}{2}-1}  \ \text{for $L$ even,} \\
 s_1^{\frac{L-3}{2}}\epsilon_1 c  \ \text{for $L$ odd,} 
\end{cases} \times \left(\frac{\pi}{\sqrt{s_{1}}}\right)^{L-1}$};

\draw (210,215) node [anchor=north west][inner sep=0.75pt]  [font=\normalsize] [align=left] {$\displaystyle =\pi ^{L-1} \times \begin{cases}
\int_1 \frac{1}{t_{1}\sqrt{s_{1}}}\ \text{for $L$ even,} \\
\int_1 \frac{c \epsilon _{1}}{t_{1} s_{1}} \ \text{for $L$ odd.}
\end{cases}$};

\end{tikzpicture}

}
}}
\end{eqnarray}
where in the first equality, we have given the part of the integrand that depends on $1$ (the white root) for even and odd $L$, and  each black leaf (a massive triangle) gives $\frac{\pi}{2\sqrt{s_1}}$ (up to overall constants); in the second equality, we are left with a one-loop integral, which gives either divergent or vanishing results from the last loop integral.

Such reductions can be applied to any graphs, and we use the following (graphic) examples to illustrate this point: 
\begin{eqnarray}
\vcenter{\hbox{\scalebox{1.0}{
\begin{tikzpicture}[x=0.75pt,y=0.75pt,yscale=-1,xscale=1]

\draw [line width=0.75]    (265.63,168.56) -- (300,168.56) ;
\draw [line width=0.75]    (146.63,168.33) -- (177.14,198.89) ;
\draw [line width=0.75]    (146.63,168.33) -- (177.14,137.47) ;
\draw [line width=0.75]    (146.63,168.33) -- (188.5,168) ;
\draw  [fill={rgb, 255:red, 255; green, 255; blue, 255 }  ,fill opacity=1 ] (173.92,137.47) .. controls (173.92,135.68) and (175.36,134.24) .. (177.14,134.24) .. controls (178.93,134.24) and (180.37,135.68) .. (180.37,137.47) .. controls (180.37,139.25) and (178.93,140.7) .. (177.14,140.7) .. controls (175.36,140.7) and (173.92,139.25) .. (173.92,137.47) -- cycle ;
\draw  [fill={rgb, 255:red, 0; green, 0; blue, 0 }  ,fill opacity=1 ] (185.27,168) .. controls (185.27,166.22) and (186.72,164.77) .. (188.5,164.77) .. controls (190.28,164.77) and (191.73,166.22) .. (191.73,168) .. controls (191.73,169.78) and (190.28,171.23) .. (188.5,171.23) .. controls (186.72,171.23) and (185.27,169.78) .. (185.27,168) -- cycle ;
\draw  [fill={rgb, 255:red, 0; green, 0; blue, 0 }  ,fill opacity=1 ] (173.92,198.89) .. controls (173.92,197.1) and (175.36,195.66) .. (177.14,195.66) .. controls (178.93,195.66) and (180.37,197.1) .. (180.37,198.89) .. controls (180.37,200.67) and (178.93,202.12) .. (177.14,202.12) .. controls (175.36,202.12) and (173.92,200.67) .. (173.92,198.89) -- cycle ;
\draw  [fill={rgb, 255:red, 255; green, 255; blue, 255 }  ,fill opacity=1 ] (143.4,168.33) .. controls (143.4,166.55) and (144.85,165.11) .. (146.63,165.11) .. controls (148.41,165.11) and (149.86,166.55) .. (149.86,168.33) .. controls (149.86,170.12) and (148.41,171.56) .. (146.63,171.56) .. controls (144.85,171.56) and (143.4,170.12) .. (143.4,168.33) -- cycle ;
\draw  [fill={rgb, 255:red, 0; green, 0; blue, 0 }  ,fill opacity=1 ] (159.42,151.97) .. controls (159.42,150.18) and (160.86,148.74) .. (162.64,148.74) .. controls (164.43,148.74) and (165.87,150.18) .. (165.87,151.97) .. controls (165.87,153.75) and (164.43,155.2) .. (162.64,155.2) .. controls (160.86,155.2) and (159.42,153.75) .. (159.42,151.97) -- cycle ;

\draw  [fill={rgb, 255:red, 255; green, 255; blue, 255 }  ,fill opacity=1 ] (265.4,168.56) .. controls (265.4,166.78) and (266.85,165.33) .. (268.63,165.33) .. controls (270.41,165.33) and (271.86,166.78) .. (271.86,168.56) .. controls (271.86,170.35) and (270.41,171.79) .. (268.63,171.79) .. controls (266.85,171.79) and (265.4,170.35) .. (265.4,168.56) -- cycle ;
\draw  [fill={rgb, 255:red, 0; green, 0; blue, 0 }  ,fill opacity=1 ] (296.37,168.56) .. controls (296.37,166.78) and (297.82,165.33) .. (299.6,165.33) .. controls (301.38,165.33) and (302.83,166.78) .. (302.83,168.56) .. controls (302.83,170.35) and (301.38,171.79) .. (299.6,171.79) .. controls (297.82,171.79) and (296.37,170.35) .. (296.37,168.56) -- cycle ;

\draw (86,151) node [anchor=north west][inner sep=0.75pt]   [align=left] {$\displaystyle \int _{1,\cdots ,5}$};
\draw (213,151) node [anchor=north west][inner sep=0.75pt]   [align=left] {$\displaystyle =\int _{1,2}$};
\draw (263.58,151.07) node [anchor=north west][inner sep=0.75pt]  [font=\footnotesize] [align=left] {$\displaystyle 1$};
\draw (315,149) node [anchor=north west][inner sep=0.75pt]   [align=left] {$\displaystyle \frac{2\epsilon_2 s_1}{c^{1/2}} \times \left(\frac{\pi ^{3}}{s_{1}\sqrt{t_{2}}}\right) =2\pi ^{3}\int _{1,2}\frac{c^{3/2} \epsilon _{2}}{t_{1} D_{1,2} s_{2}\sqrt{t_{2}}}$};
\draw (133.58,162.22) node [anchor=north west][inner sep=0.75pt]  [font=\footnotesize] [align=left] {$\displaystyle 1$};
\draw (183.04,123.15) node [anchor=north west][inner sep=0.75pt]  [font=\footnotesize] [align=left] {$\displaystyle 5$};
\draw (197.12,160.9) node [anchor=north west][inner sep=0.75pt]  [font=\footnotesize] [align=left] {$\displaystyle 3$};
\draw (180.52,199.67) node [anchor=north west][inner sep=0.75pt]  [font=\footnotesize] [align=left] {$\displaystyle 4$};
\draw (149.04,140.35) node [anchor=north west][inner sep=0.75pt]  [font=\footnotesize] [align=left] {$\displaystyle 2$};
\draw (295.38,151.07) node [anchor=north west][inner sep=0.75pt]  [font=\footnotesize] [align=left] {$\displaystyle 2$};
\draw (213,199) node [anchor=north west][inner sep=0.75pt]   [align=left] {$\displaystyle =\pi ^{4}\int _{2}\frac{c\epsilon _{2}}{s_{2} t_{2}}. 
$};

\end{tikzpicture}

}
}}
\end{eqnarray}



\begin{eqnarray}
\vcenter{\hbox{\scalebox{1.0}{
\begin{tikzpicture}[x=0.75pt,y=0.75pt,yscale=-1,xscale=1]

\draw [line width=0.75]    (146.63,168.33) -- (189,179.75) ;
\draw [line width=0.75]    (311,168.75) -- (277.63,168.56) ;
\draw [line width=0.75]    (146.63,168.33) -- (177.14,198.89) ;
\draw [line width=0.75]    (146.63,168.33) -- (177.14,137.47) ;
\draw [line width=0.75]    (162.64,151.97) -- (184.5,159.75) ;
\draw  [fill={rgb, 255:red, 255; green, 255; blue, 255 }  ,fill opacity=1 ] (173.92,137.47) .. controls (173.92,135.68) and (175.36,134.24) .. (177.14,134.24) .. controls (178.93,134.24) and (180.37,135.68) .. (180.37,137.47) .. controls (180.37,139.25) and (178.93,140.7) .. (177.14,140.7) .. controls (175.36,140.7) and (173.92,139.25) .. (173.92,137.47) -- cycle ;
\draw  [fill={rgb, 255:red, 255; green, 255; blue, 255 }  ,fill opacity=1 ] (181.27,159.75) .. controls (181.27,157.97) and (182.72,156.52) .. (184.5,156.52) .. controls (186.28,156.52) and (187.73,157.97) .. (187.73,159.75) .. controls (187.73,161.53) and (186.28,162.98) .. (184.5,162.98) .. controls (182.72,162.98) and (181.27,161.53) .. (181.27,159.75) -- cycle ;
\draw  [fill={rgb, 255:red, 255; green, 255; blue, 255 }  ,fill opacity=1 ] (173.92,198.89) .. controls (173.92,197.1) and (175.36,195.66) .. (177.14,195.66) .. controls (178.93,195.66) and (180.37,197.1) .. (180.37,198.89) .. controls (180.37,200.67) and (178.93,202.12) .. (177.14,202.12) .. controls (175.36,202.12) and (173.92,200.67) .. (173.92,198.89) -- cycle ;
\draw  [fill={rgb, 255:red, 255; green, 255; blue, 255 }  ,fill opacity=1 ] (143.4,168.33) .. controls (143.4,166.55) and (144.85,165.11) .. (146.63,165.11) .. controls (148.41,165.11) and (149.86,166.55) .. (149.86,168.33) .. controls (149.86,170.12) and (148.41,171.56) .. (146.63,171.56) .. controls (144.85,171.56) and (143.4,170.12) .. (143.4,168.33) -- cycle ;
\draw  [fill={rgb, 255:red, 0; green, 0; blue, 0 }  ,fill opacity=1 ] (159.42,151.97) .. controls (159.42,150.18) and (160.86,148.74) .. (162.64,148.74) .. controls (164.43,148.74) and (165.87,150.18) .. (165.87,151.97) .. controls (165.87,153.75) and (164.43,155.2) .. (162.64,155.2) .. controls (160.86,155.2) and (159.42,153.75) .. (159.42,151.97) -- cycle ;
\draw  [fill={rgb, 255:red, 0; green, 0; blue, 0 }  ,fill opacity=1 ] (185.77,179.75) .. controls (185.77,177.97) and (187.22,176.52) .. (189,176.52) .. controls (190.78,176.52) and (192.23,177.97) .. (192.23,179.75) .. controls (192.23,181.53) and (190.78,182.98) .. (189,182.98) .. controls (187.22,182.98) and (185.77,181.53) .. (185.77,179.75) -- cycle ;

\draw  [fill={rgb, 255:red, 255; green, 255; blue, 255 }  ,fill opacity=1 ] (272.4,168.56) .. controls (272.4,166.78) and (273.85,165.33) .. (275.63,165.33) .. controls (277.41,165.33) and (278.86,166.78) .. (278.86,168.56) .. controls (278.86,170.35) and (277.41,171.79) .. (275.63,171.79) .. controls (273.85,171.79) and (272.4,170.35) .. (272.4,168.56) -- cycle ;
\draw  [fill={rgb, 255:red, 0; green, 0; blue, 0 }  ,fill opacity=1 ] (305.27,168.75) .. controls (305.27,166.97) and (306.72,165.52) .. (308.5,165.52) .. controls (310.28,165.52) and (311.73,166.97) .. (311.73,168.75) .. controls (311.73,170.53) and (310.28,171.98) .. (308.5,171.98) .. controls (306.72,171.98) and (305.27,170.53) .. (305.27,168.75) -- cycle ;

\draw (86,151) node [anchor=north west][inner sep=0.75pt]   [align=left] {$\displaystyle \int _{1,\cdots , 6}$};
\draw (220,151) node [anchor=north west][inner sep=0.75pt]   [align=left] {$\displaystyle =\int _{1,2}$};
\draw (271,151.67) node [anchor=north west][inner sep=0.75pt]  [font=\footnotesize] [align=left] {$\displaystyle 1$};
\draw (325,151) node [anchor=north west][inner sep=0.75pt]   [align=left] {$\displaystyle s_1 t_2 \times \left(\frac{\pi^4}{s_{1} t_{2}}\right) =2\pi ^{4}\int _{1,2}\frac{c^2}{t_{1} D_{1,2} s_{2}}$.};
\draw (133.58,162.22) node [anchor=north west][inner sep=0.75pt]  [font=\footnotesize] [align=left] {$\displaystyle 1$};
\draw (183.04,123.15) node [anchor=north west][inner sep=0.75pt]  [font=\footnotesize] [align=left] {$\displaystyle 5$};
\draw (185.12,196.9) node [anchor=north west][inner sep=0.75pt]  [font=\footnotesize] [align=left] {$\displaystyle 4$};
\draw (198.02,172.17) node [anchor=north west][inner sep=0.75pt]  [font=\footnotesize] [align=left] {$\displaystyle 3$};
\draw (149.04,143) node [anchor=north west][inner sep=0.75pt]  [font=\footnotesize] [align=left] {$\displaystyle 2$};
\draw (305,151.67) node [anchor=north west][inner sep=0.75pt]  [font=\footnotesize] [align=left] {$\displaystyle 2$};
\draw (195.04,152.65) node [anchor=north west][inner sep=0.75pt]  [font=\footnotesize] [align=left] {$\displaystyle 6$};

\end{tikzpicture}

}
}}
\end{eqnarray}



\begin{eqnarray}
\vcenter{\hbox{\scalebox{1.0}{
\begin{tikzpicture}[x=0.75pt,y=0.75pt,yscale=-1,xscale=1]


\draw [line width=0.75]    (299.53,146.66) -- (277.63,168.56) ;
\draw [line width=0.75]    (277.63,168.56) -- (299.53,190.46) ;

\draw [line width=0.75]    (146.63,168.33) -- (177.14,198.89) ;
\draw [line width=0.75]    (146.63,168.33) -- (177.14,137.47) ;
\draw [line width=0.75]    (162.64,151.97) -- (181,166.5) ;
\draw  [fill={rgb, 255:red, 255; green, 255; blue, 255 }  ,fill opacity=1 ] (173.92,137.47) .. controls (173.92,135.68) and (175.36,134.24) .. (177.14,134.24) .. controls (178.93,134.24) and (180.37,135.68) .. (180.37,137.47) .. controls (180.37,139.25) and (178.93,140.7) .. (177.14,140.7) .. controls (175.36,140.7) and (173.92,139.25) .. (173.92,137.47) -- cycle ;
\draw  [fill={rgb, 255:red, 255; green, 255; blue, 255 }  ,fill opacity=1 ] (178.27,167) .. controls (178.27,165.22) and (179.72,163.77) .. (181.5,163.77) .. controls (183.28,163.77) and (184.73,165.22) .. (184.73,167) .. controls (184.73,168.78) and (183.28,170.23) .. (181.5,170.23) .. controls (179.72,170.23) and (178.27,168.78) .. (178.27,167) -- cycle ;
\draw  [fill={rgb, 255:red, 255; green, 255; blue, 255 }  ,fill opacity=1 ] (173.92,198.89) .. controls (173.92,197.1) and (175.36,195.66) .. (177.14,195.66) .. controls (178.93,195.66) and (180.37,197.1) .. (180.37,198.89) .. controls (180.37,200.67) and (178.93,202.12) .. (177.14,202.12) .. controls (175.36,202.12) and (173.92,200.67) .. (173.92,198.89) -- cycle ;
\draw  [fill={rgb, 255:red, 255; green, 255; blue, 255 }  ,fill opacity=1 ] (143.4,168.33) .. controls (143.4,166.55) and (144.85,165.11) .. (146.63,165.11) .. controls (148.41,165.11) and (149.86,166.55) .. (149.86,168.33) .. controls (149.86,170.12) and (148.41,171.56) .. (146.63,171.56) .. controls (144.85,171.56) and (143.4,170.12) .. (143.4,168.33) -- cycle ;
\draw  [fill={rgb, 255:red, 0; green, 0; blue, 0 }  ,fill opacity=1 ] (159.42,151.97) .. controls (159.42,150.18) and (160.86,148.74) .. (162.64,148.74) .. controls (164.43,148.74) and (165.87,150.18) .. (165.87,151.97) .. controls (165.87,153.75) and (164.43,155.2) .. (162.64,155.2) .. controls (160.86,155.2) and (159.42,153.75) .. (159.42,151.97) -- cycle ;
\draw  [fill={rgb, 255:red, 0; green, 0; blue, 0 }  ,fill opacity=1 ] (159.27,184) .. controls (159.27,182.22) and (160.72,180.77) .. (162.5,180.77) .. controls (164.28,180.77) and (165.73,182.22) .. (165.73,184) .. controls (165.73,185.78) and (164.28,187.23) .. (162.5,187.23) .. controls (160.72,187.23) and (159.27,185.78) .. (159.27,184) -- cycle ;

\draw  [fill={rgb, 255:red, 255; green, 255; blue, 255 }  ,fill opacity=1 ] (274.4,168.56) .. controls (274.4,166.78) and (275.85,165.33) .. (277.63,165.33) .. controls (279.41,165.33) and (280.86,166.78) .. (280.86,168.56) .. controls (280.86,170.35) and (279.41,171.79) .. (277.63,171.79) .. controls (275.85,171.79) and (274.4,170.35) .. (274.4,168.56) -- cycle ;
\draw  [fill={rgb, 255:red, 0; green, 0; blue, 0 }  ,fill opacity=1 ] (296.3,146.66) .. controls (296.3,144.88) and (297.75,143.44) .. (299.53,143.44) .. controls (301.31,143.44) and (302.76,144.88) .. (302.76,146.66) .. controls (302.76,148.45) and (301.31,149.89) .. (299.53,149.89) .. controls (297.75,149.89) and (296.3,148.45) .. (296.3,146.66) -- cycle ;
\draw  [fill={rgb, 255:red, 0; green, 0; blue, 0 }  ,fill opacity=1 ] (296.3,190.46) .. controls (296.3,188.68) and (297.75,187.23) .. (299.53,187.23) .. controls (301.31,187.23) and (302.76,188.68) .. (302.76,190.46) .. controls (302.76,192.24) and (301.31,193.69) .. (299.53,193.69) .. controls (297.75,193.69) and (296.3,192.24) .. (296.3,190.46) -- cycle ;

\draw (93,151) node [anchor=north west][inner sep=0.75pt]   [align=left] {$\displaystyle \int _{1,\cdots ,6}$};
\draw (213,151) node [anchor=north west][inner sep=0.75pt]   [align=left] {$\displaystyle =\int _{1,2,3}$};
\draw (265.58,163) node [anchor=north west][inner sep=0.75pt]  [font=\footnotesize] [align=left] {$\displaystyle 1$};
\draw (313,151) node [anchor=north west][inner sep=0.75pt]   [align=left] {$\displaystyle \frac{ \epsilon_3 t_2}{c^{1/2}} \times \left(\frac{\pi^3}{t_{2}\sqrt{t_{3}}} \right)=4\pi ^{3}\int _{1,2,3}\frac{c^{3/2}\epsilon _{1} \epsilon _{3}}{t_{1} D_{1,2} D_{1,3} s_{2} s_{3}\sqrt{t_{3}}}$};
\draw (133.58,162.22) node [anchor=north west][inner sep=0.75pt]  [font=\footnotesize] [align=left] {$\displaystyle 1$};
\draw (183.04,123.15) node [anchor=north west][inner sep=0.75pt]  [font=\footnotesize] [align=left] {$\displaystyle 4$};
\draw (151.12,185.4) node [anchor=north west][inner sep=0.75pt]  [font=\footnotesize] [align=left] {$\displaystyle 3$};
\draw (180.52,199.67) node [anchor=north west][inner sep=0.75pt]  [font=\footnotesize] [align=left] {$\displaystyle 6$};
\draw (149.04,140.35) node [anchor=north west][inner sep=0.75pt]  [font=\footnotesize] [align=left] {$\displaystyle 2$};
\draw (295,130) node [anchor=north west][inner sep=0.75pt]  [font=\footnotesize] [align=left] {$\displaystyle 2$};
\draw (213,210) node [anchor=north west][inner sep=0.75pt]   [align=left] {$\displaystyle =2\pi ^{4}\int _{1,3}\frac{c\epsilon _{1} \epsilon _{3}}{t_{1}\sqrt{s_{1}} D_{1,3} s_{3}\sqrt{t_{3}}}.$};
\draw (193.04,162.15) node [anchor=north west][inner sep=0.75pt]  [font=\footnotesize] [align=left] {$\displaystyle 5$};
\draw (295,197) node [anchor=north west][inner sep=0.75pt]  [font=\footnotesize] [align=left] {$\displaystyle 3$};

\end{tikzpicture}

}
}}
\end{eqnarray}




In these examples, we have put the factors from integrating out ``leaves" (massive triangles) in parenthesises, and for some cases this can be iterated to integrate out new ``leaves" from last step. The upshot is that these higher-loop integrals are reduced down to one-loop integral, two-loop integral for the chain graph, and another two-loop integral which equals the reduction of a four-loop chain graph, times appropriate factors of $\pi$.


\section{Discussions}
In this note we have investigated various aspects of the newly proposed four-point ABJM amplituhedron, which is obtained by reducing the kinematics to $D=3$ for the four-point amplituhedron of ${\cal N}=4$ SYM. We have initiated the study of their generalized unitarity cuts from the geometry, including the proof of perturbative unitarity and vanishing cuts from negative geometries. We have also started integrating forms of these negative geometries and obtained the infrared-finite function up to $L=3$.

These preliminary results have opened up lots of exciting possibilities in both directions considered here and beyond. It is of course important to compute/bootstrap forms of negative geometries to higher loops. The four-point ABJM amplituhedron can be viewed as a simplified model (with very rich mathematics and physics) for the one in ${\cal N}=4$ SYM, thus such computations may shed light on the SYM amplituhedron as well. Similarly one can extract infinite all-loop predictions for generalized unitarity cuts just from the geometry, and for example it would be very interesting to compute the analog of ``deepest cuts"~\cite{Arkani-Hamed:2018rsk} for ABJM four-point amplitude. On the other hand, integrating the forms to higher loops is also doable especially since the reduction identity already shows that many of them can be reduced to lower-loop ones: for $L=4$, the ``box diagram" requires more effort, but once it is done we can have $F_3(z)$ and extract from it $\Gamma_{\rm cusp}$ at next order. It is also tempting to try and find some analog of ``boxing" operator~\cite{Arkani-Hamed:2021iya}, which might allow us to resum certain bipartite graphs for any value of the coupling.  

One of the most pressing questions in this program is to nail the ABJM amplituhedron for all $n$
(with $k=\frac{n}{2}-2$ or the middle sector)~\cite{progress1}. A related approach is as follows. One can extract forward-limit {\it etc.} of higher-point amplitudes from cuts of four-point ones, which allows us to connect these geometries with different numbers of loops and legs. For example, by cutting $n=4$ integrands computed up to $L=5$, it is possible to extract $n=6$ integrands (with two possible $k=1$ leading singularities) up to $L=4$. As we have seen, such higher-point geometries can still be decomposed to negative geometries corresponding to bipartite graphs, and it would be extremely interesting to work them out already for $L=2$ cases where the integrand is known~\cite{Caron-Huot:2012sos, He:2022lfz}; based on these considerations, we expect to determine {\it e.g.} the three-loop $n=6$ integrand in terms of negative geometries. With all these fascinating structures better understood, we may gain some insights into the answer of the following question: why is there such a simple connection between (all-loop, all multiplicity amplituhedra of) ${\cal N}=4$ SYM in $D=4$ and ABJM theory in $D=3$?

\section*{Acknowledgement} We thank Yu-tin Huang for stimulating discussions and collaborations on related projects. We are grateful to Johaness Henn, Martin Lagares and Shun-Qing Zhang for sharing their results with us. This research is supported in part by National Natural Science Foundation of China under Grant No. 11935013, 12047502, 12047503, 12247103, 12225510. This work is also supported in part by the U.S. Department of Energy under contract number DE-AC02-76SF00515. C.-K. Kuo is supported by Taiwan Ministry of Science and Technology Grant No. 109-2112-M-002 -020 -MY3 as well as MOST 110-2923-M-002 -016 -MY3.

{\bf Note added}: While this work was in progress, we were made aware of an upcoming work~\cite{Henn:2023pkc}, where the finite function up to $L=3$ has been computed and the two results agree. 
\appendix

\bibliographystyle{utphys}
\bibliography{bib}

\end{document}